\documentclass[runningheads,a4paper]{llncs}
\usepackage{amssymb} 
\usepackage{graphicx}
\usepackage{url}
\usepackage{float}
\usepackage{soul,color}
\usepackage{amsmath}
\usepackage{todonotes}
\usepackage[inline]{trackchanges}
\authorrunning{ }

\begin{document}

\title{Teaching DevOps in Corporate Environments}
\subtitle{An experience report}

\author{
Manuel Mazzara,
Alexandr Naumchev,
Larisa Safina \\
Alberto Sillitti, 
Konstantin Urysov \\
\institute{Innopolis University, Russian Federation}
\{m.mazzara, a.naumchev, l.safina, a.sillitti, k.urysov\}@innopolis.ru
}

\toctitle{Lecture Notes in Computer Science}
\tocauthor{Authors' Instructions}
\maketitle
\begin{abstract}
This paper describes our experience of training a team of developers of an East-European phone service provider. The training experience was structured in two sessions of two days each conducted in different weeks with a gap of about fifteen days. The first session was dedicated to the Continuous Integration Delivery Pipeline, and the second on Agile methods. We summarize the activity, its preparation and delivery and draw some conclusions out of it on our mistakes and how future session should be addressed. 
\end{abstract}

\section{Introduction}
Our society is observing a trend of rapid technological development and a global process of automation. The public often identifies as \textit{technological progress} a new release of a telephone or of a new kind of digital device. However, technological progress is not only about ``hard'' technologies — tangible products that can be touched and ultimately purchased.
Instead, it is a balanced mixture of both technical and business innovation, including process innovation. To survive, companies must fight for every single customer, propose competitive prices, and optimize their operations \cite{BucchiaroneDDLM18}. Innovative business models appear everywhere from the game industry to the mobile application domain, and the borders of Information Technology become blurred. For example, is Uber a taxi or an IT company? Is Airbnb a realtor? The separation between Information Technology and other businesses is not so neat anymore, so that software development techniques and operations need to catch up with this trend.

It is obvious when the next release of Windows come, but what about a web service (e.g., Google, Yandex search)? Agile Methods deal with this problem only from the software development point of view focusing on customer value, managing volatile requirements, etc. However, the current needs require much more than that and involve the entire life cycle of a software system, including its operation. The DevOps approach \cite{Bass,Jabbari:2016} and microservices architectural style \cite{Dragoni2017,DragoniLLMMS17} with its domains of interests \cite{Salikhov2016a,Salikhov2016b,Nalin2016,BucchiaroneDDLM18} have the potential of changing how companies run their systems as Agile have changed how they develop their software. DevOps is a natural evolution of the Agile approaches from the software itself to the overall infrastructure and operations that is made possible by the spread of cloud-based technologies and the everything-as-a-service approaches. In this context, even the infrastructure is becoming code with all the advantages and complexity that is well-known in common software development.

However, embracing DevOps is more complex than embracing just Agile\cite{AgileDevops}. It requires changes at organization level and the development of a new skill set and approaches that need to be adopted by different teams which could be (almost) autonomous \cite{BucenaK17}. Therefore, training is of paramount importance to establish a common background for all the different groups, including the management.

Our team is specialized in delivering corporate training for management and developers. This study describes our experience of training a team of developers of an East-European phone service provider for several days. The training experience was structured in two sessions of two days each conducted in different weeks with a gap of about fifteen days in the middle in order not to disrupt for too long the working schedule. The first session was dedicated mostly to the Continuous Integration Delivery Pipeline while the second on Agile methods in general.

\section{Session I: DevOps}

The first session was conducted over two full days at the office of our customer. The training was conducted following a schedule shared in advance. Out target group was a team of developers reporting to a line manager located in a different city reachable only by flight. Previous communication with this specific team was not possible, the only information we had was partial and communicated by the remote line manager. Therefore, the original agenda had to be adjusted on site. 

\subsection{Training process}
The training covered the following topics and it was organized in four major parts, including the discussion of the survey, which appeared in several moments, although here it is for simplicity reported only at the end. Here we report the agenda and the key points of each session.

\subsubsection{Introduction}

The narrative was built as follows:

\begin{itemize}
\item Trends in IT and impact on software architectures and development.
\item Requirements volatility and Agile development.
\item Challenges of distributed development.
\item Microservices.
\end{itemize}

\begin{figure}[h]
\centering
\includegraphics[width=0.9\textwidth]{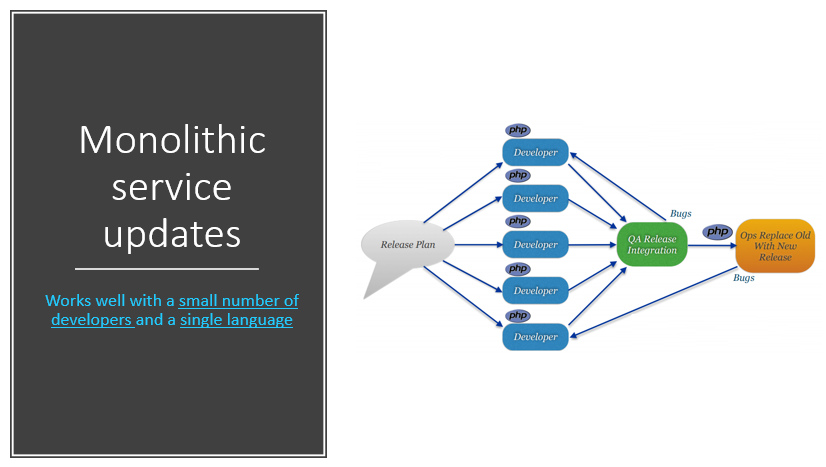}
\caption{Monolithic service updates}
\label{Monup}
\end{figure}

\begin{figure}[h]
\centering
\includegraphics[width=0.9\textwidth]{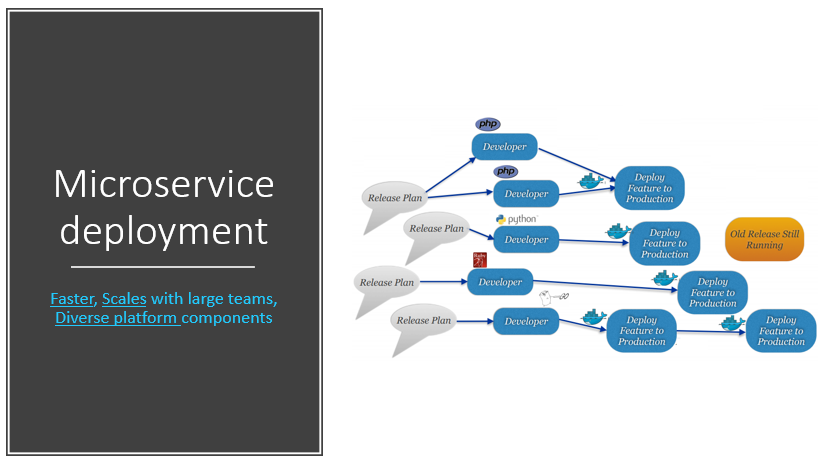}
\caption{Microservice deployment}
\label{MSdep}
\end{figure}

\paragraph{Key points.} The session emphasized the difference between \textit{hard technologies} and \textit{soft technologies}. On one side there is industrial production of commercial item of technological nature, on the other there is continuous improvement and ``agilization'' of development process. Issue of Requirements volatility and how this led to agile methods were discussed. Relevance of distributed team development in the modern setting was described in detail to conclude the session with a survey on \textit{microservices}, which are considered the privilege architecture for DevOps with their scalability features \cite{DragoniLLMMS17}. The key difference between monolithic service updates Fig.\ref{Monup}) and microservice deployment (Fig.\ref{MSdep}) was presented to motivate the need of migration to microservices and why this topic is so closely related to DevOps.
Fig.\ref{Monup} shows that the approach works well with a few developers and a single language, and that the need for microservices emerges for large teams and diverse platform components.

\subsubsection{Continuous Integration Delivery Pipeline}

Here the full delivery pipeline was discussed:

\begin{itemize}
\item Source code control.
\item Build automation.
\item Automated testing. 
\item Static code analysis.
\item Integration testing.
\item Deployment automation.
\item Monitoring.
\end{itemize}

\paragraph{Key points.}
The overall session worked on the idea that it is desirable that software in mainline can be rapidly built and deployed to production at any point. The declared benefits of this approach are:

\begin{itemize}
\item Reduction of manual effort.
\item Acceleration of release cycles.
\item Improvement of release quality.
\item Increased collaboration between development, QA, support and operations teams.
\item Reduction in costs for deployment and support.
\end{itemize}

The fundamental principles of DevOps as generally agreed upon by the most influential early members of the DevOps community, were summed up in the acronym ``CAMS'': \textbf{C}ulture, \textbf{A}utomation, \textbf{M}easurement, \textbf{S}haring.

\subsubsection{Tools}

We analyzed tools for the following purposes:

\begin{itemize}
\item Version control.
\item Build automation.
\item Testing (including mutation testing \cite{mutants}).
\item Continuous integration.
\item Configuration management.
\item Continuous monitoring.
\item Seamless development (\cite{Meyer:1997:OSC:261119}, \cite{walden1995seamless}, \cite{Meyer13Multi} \cite{Naumchev2016UnifyingExample}, \cite{Naumchev2016CompleteDrivers}, \cite{Naumchev2017}).
\end{itemize}

\paragraph{Key-points.}
This part emphasized on the practical aspects showing tools to support the idea of continuous delivery. During the discussion with teams, we were asked to speak about the mutation testing, which was out of scope of our main topic. However, we found it interesting enough to adapt our initial agenda and cover it. We have also decided on a little experiment and included the topic on seamless development, to see how well ideas born in academia can be spread among the industry.

\subsubsection{Discussion of the survey}

The survey data collected before the training was analyzed question by question to give focused and specific advice to the team, apart from the generalities discusses in the previous parts.

\subsection{Objectives of the training}

The idea of the training was not just to lecture the theory behind DevOps, but also to address specific software process-related issues preventing the client to benefit from applying DevOps as much as possible. To figure out these issues, we had to collect some data about the audience. Due to the high level of corporate privacy, we could not use internal data of the company. The schedule was tight, which is why we could not afford negotiating possible non-disclosure agreement that would let us gain the insights we needed. We decided to develop of a questionnaire that, in our opinion, would uncover some insights about the team. The questionnaire consisted of 17 questions related to DevOps practices, and in the end, coincidentally, we have received 17 responses.

\begin{figure}
\includegraphics[scale=0.7]{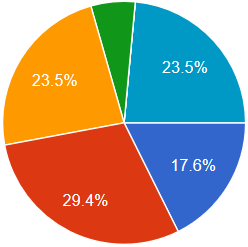}
\centering
\caption{How much time does it typically take to deploy changes? (Blue for ``less than one hour'', light blue for ``don't know'', red for ``less than one day'', yellow for ``one day to one week'', green for ``one week to one month''.)}
\label{fig:q5}
\end{figure}

\begin{figure}
\includegraphics[scale=0.7]{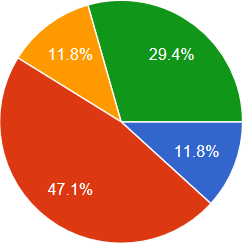}
\centering
\caption{Do you automate application testing? (Blue for ``yes'', red for ``mostly yes'', yellow for ``no'', green for ``don't know''.)}
\label{fig:q9}
\end{figure}

\begin{figure}
\includegraphics[scale=0.7]{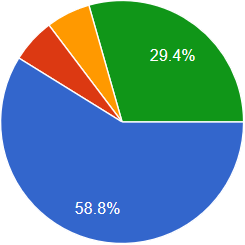}
\centering
\caption{Does your team practice retrospective and postmortem meetings? (Blue for ``yes'', red for ``only retrospectives'', yellow for ``only postmortems'', green for ``no''.)}
\label{fig:q14}
\end{figure}

\subsection{Analysis of the results}
Although we were expecting to understand something about the current level of DevOps practices in the team, the responses to the questionnaire made us think about communications inside the team. Running the questionnaire and analyzing the results revealed the following anomalies in the structure of the responses:
\begin{itemize}
\item High diversity of opinions for some questions (Fig. \ref{fig:q5}).
\item A lot of ``don't know'' replies for some straightforward questions(Fig. \ref{fig:q9}).
\item Contradicting responses for some questions (Fig. \ref{fig:q14}).
\end{itemize}



These anomalies raised the following questions to address during the training:

\begin{itemize}
\item Does the team consist, in fact, of several sub-teams?
\item To what extent are the teams distributed?
\item How does the level of DevOps practices' maturity vary among the teams?
\item How should we adjust the contents of the training to meet the identified variability?
\item How should we adjust the questionnaire itself, so that we minimize the likelihood of the mentioned anomalies in the future?
\end{itemize}

An ideal solution would be to create another questionnaire targeting these questions.
The schedule, however, did not allow us to do so.
The only solution was to figure out the missing information on site, and then adjust the training on the go, between the two training days.
By the end of the first day of the training, we have revealed the following information, with respect to the above questions:
\begin{itemize}
\item The team consists of 4 sub-teams. Each sub-team has their own goals, problems, and concerns about the software process.
\item The smallest team, consisting of 4 members, is from another city. Some teams are distributed.
\item The smallest team is not even considering applying DevOps because of its size. One of the sub-teams possesses good understanding of DevOps and applies it in their daily practices.
\item One team’s customer is the head branch of the company.
\item Every day the customer tells the team what to deploy during the day.
\item Fridays have the largest number of deployments requested.
\item Around 80\% of releases are tested after deployment.
\item Sometimes they work on weekends, which is a natural consequence of the previous two points.
\item The customer does not want the team to use automatic deployments, because they lack trust to the team.
\end{itemize}

These important pieces of information have little to do with the questionnaire that we sent to the trainees. We will use this information when it comes to composing a questionnaire for another training.

After having the above points figured out during the first day of the training, we have spent a half of the night before the second day to completely rework the contents.

\section{Session II: Agile}
The second session was held for two full days in the same office. At the beginning, the main topic was ``Agile software development'' and Scrum in particular. However, the requirements were changed at the last moment, as the customer asked to pay more attention on Kanban, since they were thinking to use it in the future. Overall, it was clear that development team was in doubt which process they need to follow.

\subsection{Training objectives and process}
\subsubsection{Gathering more data}
Due to changing requirements from the company it was clear that one of the most important demand for this training will be identification of suitable methodology for development teams. 
To achieve this goal, we decided to use a framework described in \textit{``Choose your weapon wisely''} \cite{Rockwood}. It provides information about different trade-offs for popular development processes:
\begin{itemize}
\item Rational Unified Process (RUP) \cite{rupe}
\item Microsoft's Sync-and-Stabilize Process (MSS) \cite{Cusumano:1997:MBS:255656.255698}
\item Team Software Process (TSP) \cite{Humphrey:1999:ITS:1408380}
\item Extreme Programming (XP) \cite{xp}
\item Scrum \cite{Schw01a}
\end{itemize}

\begin{figure}[h]
\centering
\includegraphics[scale=0.4]{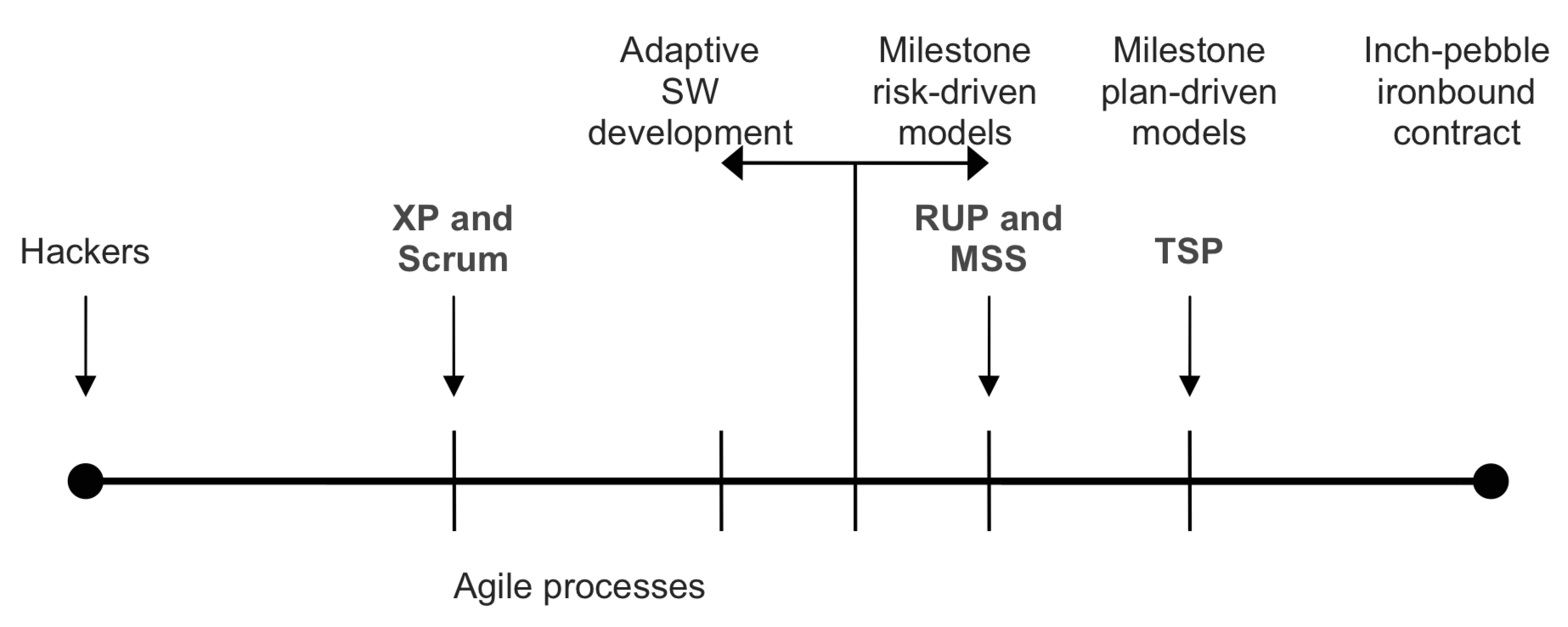}
\caption{Spectrum of processes (from Rookwood \cite{Rockwood})}
\label{fig:spect}
\end{figure}

Information about processes was divided in four parts:
\begin{enumerate}
\item Overview (short description of the process).
\item Roles (information about positions for the process).
\item Artifacts to produce (including documentation).
\item Tool support (overview of tools available on the market for using the process).
\end{enumerate}

\subsubsection{Picking the right process}
\begin{figure}[h]
\centering
\includegraphics[scale=0.35]{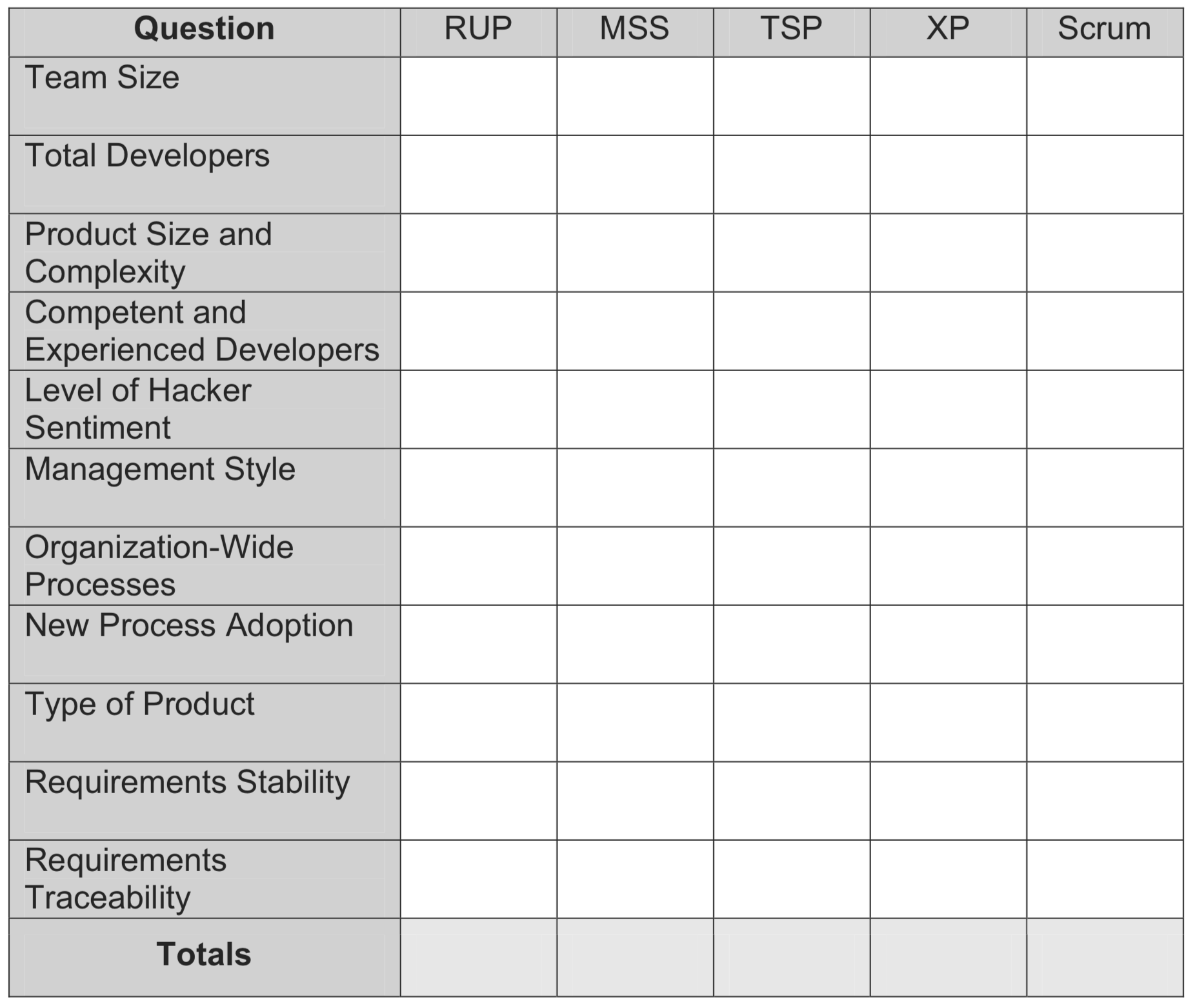}
\caption{Question Tally Sheet (from Rookwood \cite{Rockwood})}
\label{fig:MSdep}
\end{figure}

After a brief introduction of all processes, the development teams were asked to fill out the forms with a set of questions (see picture \ref{fig:MSdep}). All the questions were divided on four main groups:
\begin{itemize}
\item Team and product size (number of engineers involved in a single development team and product size in terms of LOC and complexity).
\item Developers and organization (number of competent and experienced engineers). 
\item Product and its types (life critical, embedded, ERP system, etc.).
\item Requirements and their stability.
\end{itemize}

Three teams participated in a survey, as a result it was identified that Agile methodologies (XP and Scrum) fit better than others for 90\% of respondents. Thus, it was decided to focus on Agile development processes — Scrum, XP and Kanban.

\subsection{Analysis of the results}
There is a huge difference in teaching students and mature engineers. The last ones already have prior knowledge and their own opinion on how things should be done. There is a common pattern in how teams are changing their development processes.

\subsubsection{Phase one. Waterfall}
One of the biggest problems of waterfall, besides those problems of working with emerging requirements, are \textit{``walls''} between different departments inside the organization. The Waterfall model typically does not imply cross-functional teams, meaning that different teams work independently from each other on different parts of the project. In such organizations it is typical to see parts of a deliverable moving between this institutional \textit{``walls''}, from 
department to department. For example, the analytics team collects the requirements, provide the specification and then pass it over to developers; developers code it as fast as possible and pass it to testers, who have most of the problems. The most obvious drawbacks of this approach is that testers receive one large piece of functionality without having any strong support from developers. 

\subsubsection{Phase two. Water-Scrum-Fall}
After a while the team will identify obvious problems of waterfall model and will try to change development processes, as agile methodologies very popular nowadays there is a big chance that the team will choose it. However, changing the process in big organization is complicated, and as a first result of that changes we can get water-scrum-fall model, where the whole team will be split on three parts: managers, development and QA teams and customer(s). This separation breaks several important rules of agile approach. For example, estimations quite always produced by managers based on value and money without interaction with development team, development team do not have access to the customer, which eventually adds more bugs and change requests.

\subsubsection{Phase three. Tailored Scrum with practices from XP and Kanban}
The hardest part of changing development process is a move from second to the third phase, and the key of success here is a strong leader of development team, who will manage these changes and will be able to convince all team members to follow picked processes.

\section{Lesson learned and conclusions} 
Our experience thought us a few relevant lessons:
\begin{itemize}
\item It is important to get in advance as much information as possible about the audience the training meant to.
\item Talking directly to the relevant people, possibly technical and on-site.
\item Be clear on the outcome of the training, be sure that what the audience need and expect corresponds to what we want to present.
\end{itemize}

In our case we had limited access to the development team prior to the training. Previous Skype calls only happened with management residing in a different city. We were only able to collect information via a generic questionnaire on development practices. We realized that the questionnaire may have been ambiguous in some parts, and possibly too generic. For example, we were not prepared for the fact that more than one heterogeneous team would be participating, which made some questions irrelevant and did not target all the participants. 

Therefore, we had to spend considerable amount of time collecting missing information on-site about the teams participating and technologies they used. At the end of the first day we decided to refocus our program, and the second day was mostly related to QA practices which targeted the investigated problems better.

This case was a reminder for us of utterly importance of collecting requirements and how things can be easily misunderstood or slow down if there are any obstacles during this process. Both refocusing the topic during the DevOps session and changing the topic for the following Agile session led us to the conclusion that it is not always easy to collect all required information beforehand, especially with limited ways and amount of time, and this information may not always show the real problems teams are struggling with. So, the trainers themselves should be ready to step aside from the main topic.

\bibliographystyle{abbrv}
\bibliography{biblio}

\end{document}